# Analysis of the Patent of a Protective Cover for Vertical-Axis Wind Turbines (VAWTs): Simulations of Wind Flow

**Moleón Baca, J.A. *[,1], Expósito González, A.J. [2] and Gutiérrez-Montes, C. [2]**

[1] Department of Physics, University of Jaén (Spain), Campus Universitario de las Lagunillas s/n, edificio A-3, 23071 Jaén, Spain; jamoleon@ujaen.es

[2] Fluid Mechanics Division, Department of Mechanical and Mining Engineering, University of Jaén (Spain), Campus Universitario de las Lagunillas s/n, edificio A-3, 23071 Jaén, Spain; ajexgon@gmail.com (E.G.A.J.); cgmontes@ujaen.es (G.M.C.)

**Abstract:** This paper presents a numerical and experimental analysis of the patent of a device to be used in vertical-axis wind turbines (VAWTs) under extreme wind conditions [1]. The device consists of two hemispheres interconnected by a set of conveniently implemented variable section ducts through which the wind circulates to the blades. Furthermore, the design of the cross-section of the ducts allows the control of the wind speed inside the device. These ducts are intended to work as diffusers or nozzles, depending on the needs of the installation site. Simulations were performed for the case of high-speed external wind, for which the ducts act as diffusers to reduce wind speed and maintain a well-functioning internal turbine. Four different patent designs were analyzed, focusing on turbine performance and generated power. The results indicate that the patent allows the generation of electric power for a greater range of wind speeds than with a normal wind turbine. The results support that this patent may be a good alternative for wind power generation in geographic areas with extreme weather conditions or with maintained or strong gusty wind. Experimental tests were carried out on the movement of the blades using the available model. Finally, the power curve of the model of this wind turbine was obtained.

## 1. Introduction

In the current state of the global energy market, there is a great need to continue developing renewable energy, thus reducing dependence on fossil fuels and avoiding emissions of greenhouse gases, as well as to benefit from the other advantages of this type of energy. Wind power is the source of renewable energy with the greatest potential due to its high production capacity. This source of energy is currently used primarily to produce electricity through large horizontal-axis wind turbines, which are grouped in parks and connected to the distribution grid. On the other hand, small wind turbines have been developed for use in remote locations or those isolated from the electrical grid, similar to the use of photovoltaic solar panels.

In a hundred countries around the world, wind energy is increasingly used, with a total installed capacity that far exceeds half a million megawatts; China leads the growth with around 20 GW installed in recent years [2]. In some countries such as Denmark, electricity from wind power exceeds 25% of total consumption. In Spain, it surpasses nuclear energy, reaching peaks of 40% of the energy consumed, specifically on windy days [3]. In addition, this type of energy has a less damaging impact on the environment than other renewable sources of energy. Indeed, in recent years, wind technology has had the highest rate of production out of all new installations of renewable electrical power.

The advantages of using wind energy for electricity production are clear, but there are also significant drawbacks. In systems within the network, the wind speed and energy produced have to be predicted in advance. If the wind speed is very high (more than 50 km/h), the blades must be stopped for safety reasons. Another disadvantage is that each tower requires a large space without any obstacles around it. This also applies to small generators in urban areas, which also need space so that the rotation of the blades is not a danger to people. In addition, the disturbing noise produced by the rotation of the blades limits the use of wind energy in residential areas. There are also some bird mortalities caused by collision with the blades in motion [4,5].

Furthermore, in geographical areas with permanent or occasional extreme conditions of wind or temperature, the use of renewable electric energy supply is very limited, for example, in sites that experience hurricanes or typhoons and populations or facilities near the polar circles. For a long time, research stations in Antarctica have been trying to reduce fossil fuel consumption with the use of renewable energy because these facilities have a special ecological sensitivity. Photovoltaic panels and wind turbines have special technical and installation requirements in order to be installed in this place due to the existence of strong gusts of wind, very low temperatures, and accumulation of snow [6,7].

The devices used for wind generation have been considerably developed in a relatively short time, and there have been noticeable improvements in design and performance [8–10]. Horizontal-axis wind turbines (HAWTs) are the most widespread due to their higher efficiency and production rate. This is because the blades can be very long and, therefore, occupy a larger area facing the wind. However, due to the disadvantages of HAWTs, vertical-axis wind turbines (VAWTs) have been developed for specific applications due to their advantages, such as not needing orientation systems depending on the wind direction, low starting torque, simplicity, and low cost [11–16]. Furthermore, there are many different proposals to improve the efficiency of these systems, such as the use of new design channels, ejectors, and upstream deflectors, as well as different blade configurations. Within these proposals, there are many experimental [17–24] and theoretical or numerical [25–36] investigations. These works show the interest that exists for these types of wind turbines, and they open up a wide range of possibilities for actual use [37] in large, medium, and small electricity generation facilities. Another case is the use of diffuser-augmented wind turbines (DAWTs), which has been extended in recent years in order to increase the performance of wind turbines [38,39]. Most of the devices that implement turbines focus on HAWTs [40] and, to a lesser extent, on VAWTs [41].

However, as has been previously noted, the use of these kinds of systems is not feasible in locations with extreme winds unless wind-tenable structural and productive conditions can be provided. In this regard, there is a low number of patents related to covers for wind turbines, and none are specified to work with strong winds.

In that regard, the Utility Model ES1057020U consists of a wind turbine on a roof with a corresponding reel-turbine and electric generator. This model is characterized by having parabolic blades with a ladle-like form without any rear exit, supported by a lower base and a flat cover. The rotation device (reel-turbine) is contained in a box with walls that channel both the input and output of the wind flow, regardless of the wind direction. These walls allow the rotation speed of the reel-turbine to be increased; therefore, this device is designed only to take advantage of light winds, just like the following.

Patent ES2182699B1 is a structure formed by a reinforced concrete body with a circular or polygonal section. On the inside, there are pipelines that gradually reduce its perimeter, causing the wind to have a tangential impact on the blades of the central rotor of the vertical axis. The body of the structure is protected by a metal shutter that can be regulated electronically, which is placed around its lateral surface. The base and top of the structure are inclined. The patent describes a body with pipelines, an overhang, and a sloping roof, except in the central area of the rotor.

In addition, two other Chinese patents refer to VAWT covers: (1) a vertical turbine with a wind-collecting cover (CN102094753 (A) and (B)) that lets the wind pass up to eight blades, which are specially arranged to take advantage of its strength, to improve efficiency; (2) CN201474853(U) is a VAWT with a wind-shielding cover, which has a special opening and four concave blades (that is, a Savonius type), that is also used for weak winds.

At the time of this study, there was no existing device available for application under extreme climatological conditions. Recently, a cover for VAWTs was patented (ES2482872 A1 and B1) [1]. This patent has two possible applications for the production of electrical energy. On the one hand, it can be used in places with strong, sustained, or gusty winds. In this case, the ducts are designed as diffusers to decrease wind speed when necessary. On the other hand, this system can be used to potentiate weak winds. Here, the ducts are designed as nozzles. In this way, the protective cover allows the use of domestic or medium wind power installations where the wind has low speeds.

Through a numerical study, the paper analyzes the new design of this cover in order to determine whether it is possible to make use of or even improve the performance of a wind turbine for such conditions. Thus, the possibilities of its installation and use will be discussed, but only for the first case (ducts as diffusers) because this is a novel use that has seldom been seen so far.

Therefore, the idea of this patent arose in order to try to provide a kind of renewable energy installation for extreme weather conditions, which is why most of the simulations were performed with strong winds. Thus, in Section 2, the device is fully described in detail. In Section 3, the materials, methods, and numerical model used to perform the simulations are presented. The results of the simulation of air movement inside and outside the presented design, as well as of the experiments that were carried out, are discussed in Section 4. Finally, Section 5 is devoted to the conclusions.

## 2. Description

The patented device [1] analyzed in this paper is a protective cover for vertical-axis wind turbines. This cover consists of two coaxial hemispheres of different diameters connected by ducts installed through which the wind circulates to the blades. Figure 1 shows the plans, and Figure 2 shows photographs of a model of the project. The ducts are designed so that, regardless of the wind direction, the wind passes through one or two of them and impinges on the blades of the generator so that the blades always rotate in the same direction. Then, the wind exits out of the remaining ducts and air vents at the base. According to the patent description, the number of blades and channels may vary. In the present work, the best configurations will be explored for the specific case at hand, although these could vary depending on the wind conditions. The outer dome is expected to withstand hurricane winds, and the external openings of the ducts can be built with shutters that can be closed if necessary.

The present design of a protective cover for VAWTs has some advantages over conventional wind turbines.

One of the first issues to consider is that the size of the two domes and the wind turbine is not limited. This means that small domes can be built from about one meter in diameter up to several meters. On the one hand, the latter allows a power supply using renewable energy in isolated areas or protected places for small scientific, domestic, or agricultural facilities. It also enables the production of energy in urban buildings or areas subject to extreme weather conditions, such as near the polar circles. On the other hand, this also allows the installation of large wind farms connected to the network in places previously not feasible, such as those subject to periodic hurricanes or typhoons.

Another advantage of this patent is the design of the ducts for wind. Cross-sections may be varied to regulate the wind speed (according to the continuity equation of fluid motion). This allows the wind turbine to operate with a wider range of wind speeds, both weak and strong winds. The ducts are designed as diffusers to decrease the wind speed and, conversely, as nozzles. It is expected that these features will improve the standard power curve. The number and size of the diffusers will be discussed in this paper. In addition, the positions of the ducts allow the VAWT to function regardless of the wind direction and without the need for any guidance system.

The outer dome shields the rotating blades and prevents accidents to people or animals that may be close by. In this way, it is possible to install small- or medium-sized wind turbines for domestic use, which was previously problematic. It will also reduce protests over the high rate of bird mortality due to wind turbines.

The protective cover, together with the wind turbine, is placed on a base with adequate capacity to contain the other elements necessary for the generation of electricity and its transmission or storage. This base may be underground or elevated as appropriate. The manufacturing materials can be of different types according to their location or operating conditions. Very resistant materials can be used, such as metals or lighter materials such as plastic or polyurethane resins.

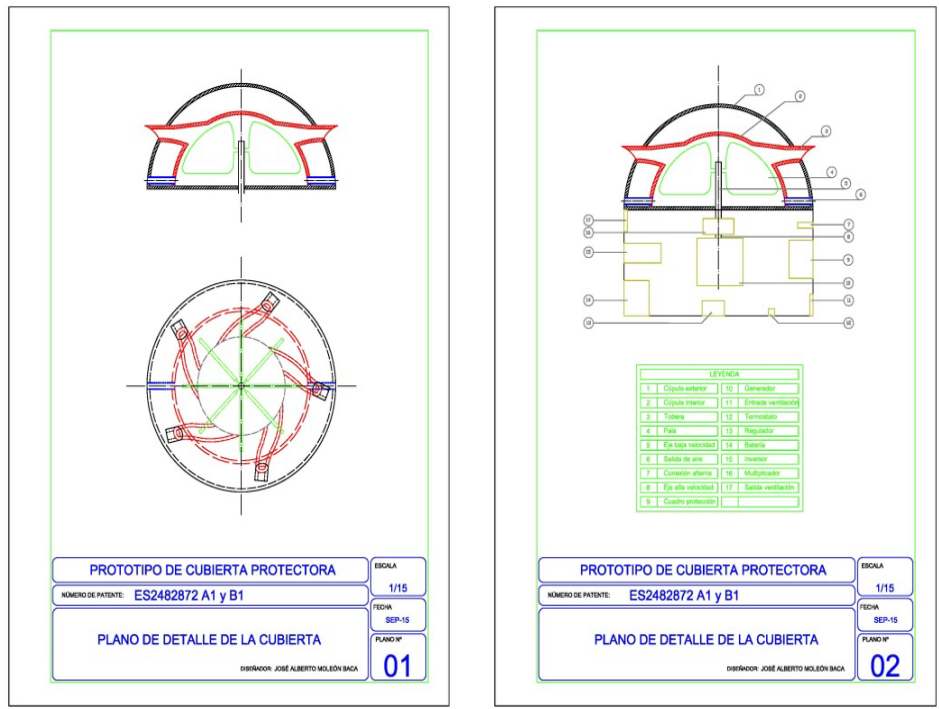

**Figure 1.** Initial plans of patented project ES2482872 A1 and B1.

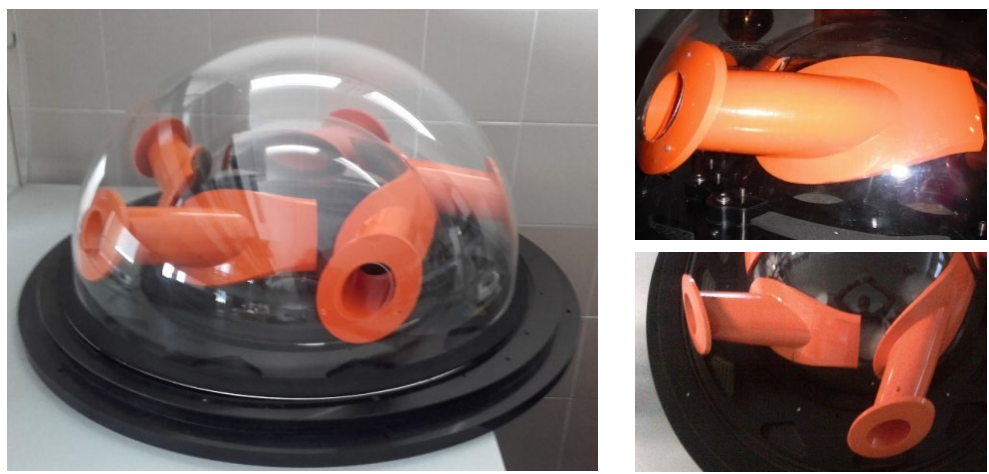

**Figure 2.** Photographs of the model of the patented project and details of the ducts (diffusers).

## 3. Materials and Methods

For this work, a model of the patent project has been manufactured in the Laboratory of Mechatronics and Robotics (LMR) of the University of Granada. The hemispheres are made of transparent methacrylate, the diameters of the external and inner domes being equal to 105 and 80 cm, respectively, and their thickness equal to 1 mm. The ducts and concave blades were manufactured with a 3D printer and are made of plastic. To conduct an experimental study on the blades, two axes were built of aluminum. In this regard, one of the axes had symmetrical supports for three or six blades, and the other for four or eight.

The controlled wind tests were carried out in a wind tunnel that allows a maximum test air velocity of 40 m/s. The rotational speeds of the rotor were measured with time signals from photoelectric cells. The measurements were performed for increasing wind speeds up to 30 m/s and different configurations of the axis and blades.

*3.1. Numerical Model*

The three-dimensional unsteady turbulent flow of an incompressible free-stream of velocity $U$, density $\rho$, and viscosity $\mu$, around and inside the semispherical-shaped protective cover, is considered herein, where the fluid is air and the free stream velocity is equal to 100 km/h. Figure 3 shows a sketch of the problem and the computational domain, $\Omega$. The body presents an internal diameter (d = 80 cm), with separation distance (h = 25 cm) between the inner and outer walls. The total dimensions of the computational domain are L = 10 (d + h), W = 10 (d + h), and H = 6.5 (d + h).

As will be later shown, different 3D geometries were designed using the CAD software SolidWorks® [42–44]. Furthermore, the computational grid was built using the ICEM CFD [45] meshing package from ANSYS INC v16.2. The grid consists of an unstructured tetrahedral mesh in which a compromise between cell size and computational cost has been pursued. A special focus was paid to the grid resolution at the regions beneath the cover walls (see Figure 3). Thus, the resolution was significantly enhanced to properly predict the flow behavior in these regions. The initial meshing was conducted by means of the Octree method, and then the mesh was refined and its quality improved through a Delaunay triangulation method. Furthermore, the proper quality of the meshes was assured through the corresponding parameters (quality index > 0.3, minimum angle > 18°, skewness < 0.85, and aspect ratio <100). Finally, FLUENT software [46] also from ANSYS INC was used to solve the governing equations. In this regard, the incompressible, turbulent, unsteady, three-dimensional flow simulations were performed using a RANS (Reynolds-averaged Navier–Stokes equations) model, namely, the $k$-$\varepsilon$ standard model and an SST model. Second-order upwind and Green–Gauss linear centered discretization schemes were used for the spatial terms and a second-order scheme for the time integration, while the pressure–velocity coupling was accounted for through the pressure-implicit split-operator (PISO) algorithm [47]. Appropriate boundary conditions were used, namely, a uniform fluid stream of velocity ($U$; 0; 0), with a turbulent intensity of 5%, and a characteristic turbulent length scale of 0.1 $H$ were imposed at the inlet, whereas an outflow boundary condition was applied at the outlet. Furthermore, a no-slip condition was set at the remaining wall boundaries, i.e., at the inner and outer walls of the protective cover, while slip conditions were imposed at the side, top, and bottom boundaries. In addition, a grid sensitivity study was carried out for four different meshes, focusing on the characteristic average drag coefficient $\overline{C_D} = \overline{F_x}/(0.5\rho U^2 (d/2 + h)^2)$ (see Table 1). As an outcome, a mesh of 897,778 cells, i.e., mesh #3, was used to perform the simulations.

*3.2. Validation*

The numerical model was subsequently compared against preliminary experimental measurements for different flow velocities at selected points of the original Model 1 for validation purposes. A hot-wire anemometer was used to measure the fluctuations of the streamwise component of the velocity at the inlet and outlet sections of the frontal duct. The temperature was monitored to correctly calculate the density and viscosity of the air stream. Moreover, the model was carefully aligned with the free-stream to match the wind direction relative to the geometry in the simulations. Table 2 shows the velocity results for the numerical calculations and the experiments, for which overall good agreement was found with relative errors below 10%, thus, corroborating the validity of the numerical model implemented herein.

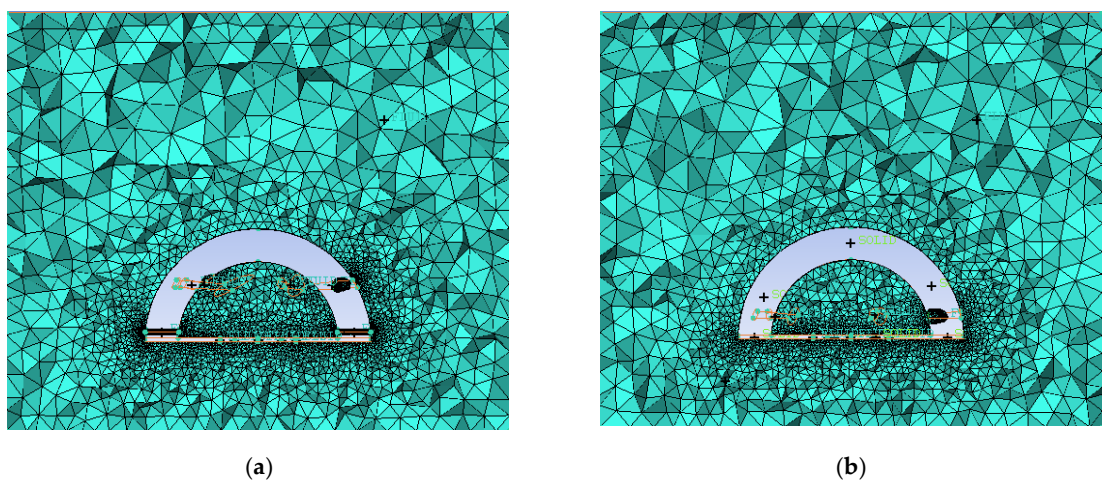

(**a**) (**b**)

**Figure 3.** The meshes used in Models 1 (**a**), Models 4 (**b**).

**Table 1.** $\overline{C_D}$ and relative errors obtained for different grids. Grid #3 was used in the simulations.

|        | Cells     | $\overline{C_D}$ | Relative Error (%) |
|--------|-----------|--------|---------------------|
| Mesh #1 | 205,946   | 0.4599 | 1.75                |
| Mesh #2 | 589,783   | 0.4462 | 1.27                |
| Mesh #3 | 897,778   | 0.4496 | 0.52                |
| Mesh #4 | 1,340,477 | 0.4512 | -                   |

**Table 2.** Experimental and numerical streamwise velocity at the inlet and outlet of the frontal duct in Model 1.

|  |  | Duct Velocity (m/s) | | | | | |
|--|--|--|--|--|--|--|--|
|  |  | Inlet | | | Outlet | | |
|  |  | Exp | CFD | Relative Error (%) | Exp | CFD | Relative Error (%) |
| Free-stream velocity (m/s) | 15 | 18.1 | 17.0 | 6.1 | 17.6 | 16.6 | 5.7 |
|  | 20 | 23,7 | 22.3 | 5.9 | 23.4 | 21.1 | 9.8 |
|  | 28 | 34.3 | 32.4 | 5.5 | 34.1 | 31.9 | 6.5 |

## 4. Results

In the present work, special focus was paid to diffuser-type ducts due to their more innovative use and usefulness. Thus, numerical simulations of different 3D models were conducted, and the inlet conditions were analyzed. The results for the simplest model with five straight diffusers are presented first, followed by those corresponding to alternative models that have yielded better results (Figure 4). Wind turbine blades, as seen in the figures, were not included in the simulation but have been left in the model for the sake of clarity as they indicate how important the design of the ducts is to maximize the torque on the rotor to improve its performance. The blades were later included in the experimental part that is discussed below.

Model 1 is a simple design that serves as a starting point, with five straight ducts leading the wind to the turbine and two additional tubes at the bottom for air release. As will be later observed, these two tubes are unnecessary because there are already enough air outlets at the base of the dome. In Model 2, a curvature is introduced to the diffusers, with a consequent variation in the input and output cross-sections. Furthermore, as will be later shown, the ducts' layout in the design of Models 3 and 4 was modified, based on the results of the first simulations, to gain from the dynamic pressure of the wind and to also direct the wind to the lower part of the blades, optimizing the efficiency of the device. Following these changes, it was observed that the lengths of the diffusers were also shortened. Finally, the number of diffusers was increased from five in Model 3 to eight in Model 4 in order to enlarge the wind inlet area and, thus, the power generated (Equation (1)).

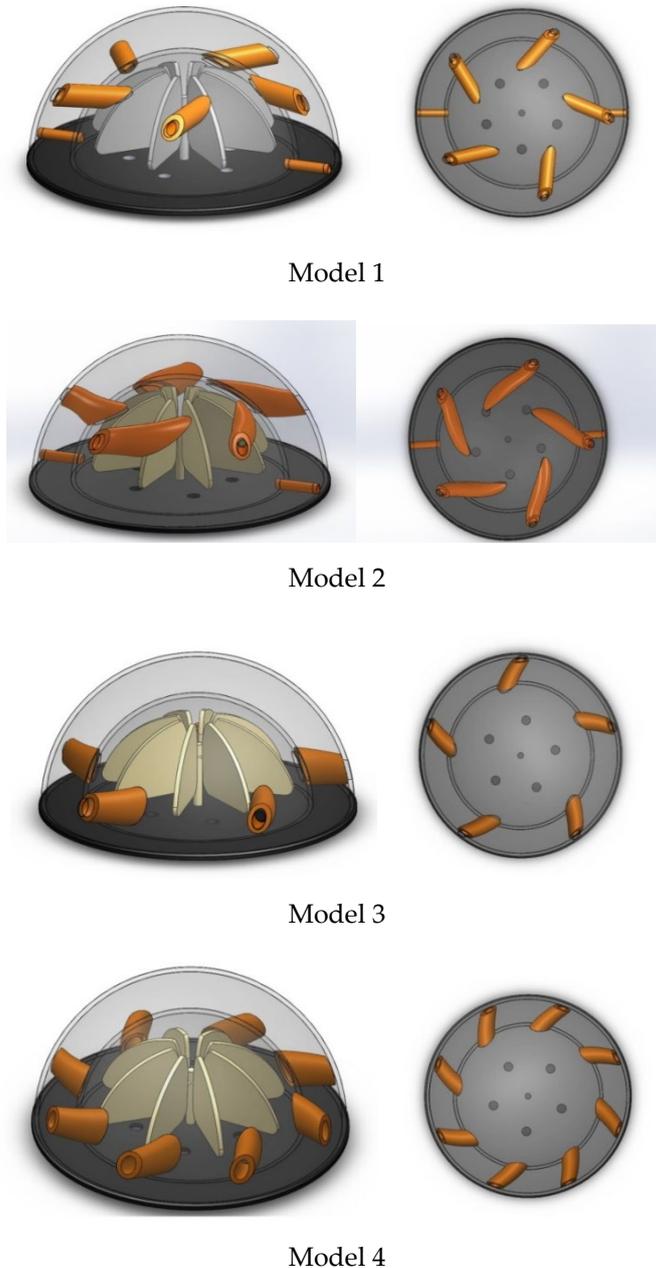

Model 1

Model 2

Model 3

Model 4

**Figure 4.** Wind turbine models used for the numerical simulations.

*4.1. Simulation Results*

Numerical simulations of the aforementioned four models were conducted. Table 3 shows the main dimensions of the diffusers. Specifically, the input and output areas correspond to the size of the openings in the outer and inner domes, respectively. The diameters of the external and inner domes are 105 and 80 cm. The input area is the size of each hole of the outer dome, through which the air enters the inner dome. The output area is the size of each hole of the duct discharging into the inner dome.

**Table 3.** Diffusers sizes in each model.

| Diffusers | Input Area (mm$^2$) | Output Area (mm$^2$) | Volume (mm$^3$) |
| --- | --- | --- | --- |
| Model 1 | 28.18 | 43.48 | 655.48 |
| Model 2 | 30.43 | 180.69 | 1583.75 |
| Model 3 | 26.98 | 69.01 | 814.83 |
| Model 4 | 26.98 | 69.01 | 814.83 |

The results of the simulations are presented next. To that aim, the contours of pressure and velocity, as well as the velocity vectors field, are shown for the models studied herein. In the simulations, the inlet wind velocity was set to 100 km/h (27.78 m/s) to determine how it is reduced inside the cover in order to make it suitable for the normal conditions of standard wind turbines in operation. Finally, the power curve of the device for different wind velocities will be discussed.

First, the pressure contours (Pascals) are considered. Figure 5 shows the stagnation region with the largest static pressure, which can be clearly identified. This region is located below the inlet openings of the diffusers of Model 1. When the fluid reaches the obstacle, it slows down and thus increases its pressure, transforming the dynamic pressure into static pressure. Moreover, there are areas of low pressure, which correspond to the regions where the flow accelerates on top of the protective cover, and two others that are symmetric on the sides, where no flow detachment occurs. It is also worth noting that there is also no flow separation at the rear part where the wind exits.

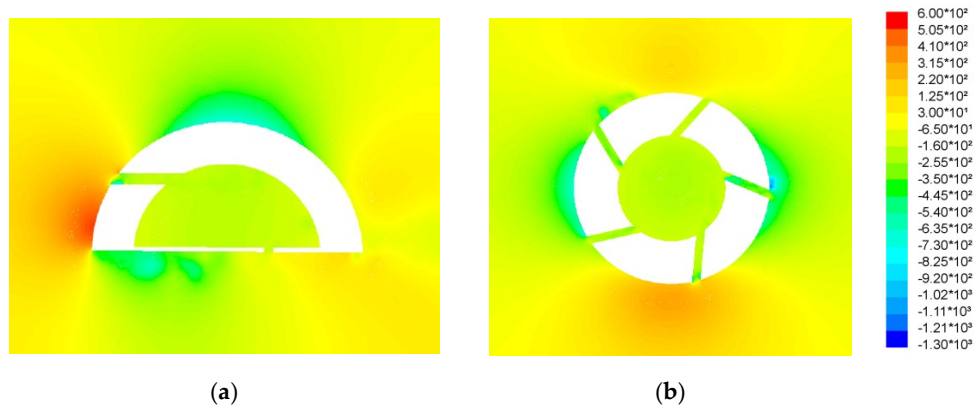

(**a**) (**b**)

**Figure 5.** The pressure contours (Pascals) of Model 1: (**a**) on a longitudinal cutting to the diffusers; (**b**) on a transverse plane in the diffusers.

Figure 6 shows the velocity contours (m/s) for Model 2. Again, the stagnation area (in blue) can be clearly identified; it is characterized by low speed but high pressure. At this point, the flow accelerates as it enters the diffuser (reddish tone) due to an abrupt geometric change when the wind reaches the opening. This leads to a vein contraction phenomenon, which occurs through recirculation at the bottom of the diffuser due to a decrease in its diameter, and consequently, an increase in the velocity of the flow.

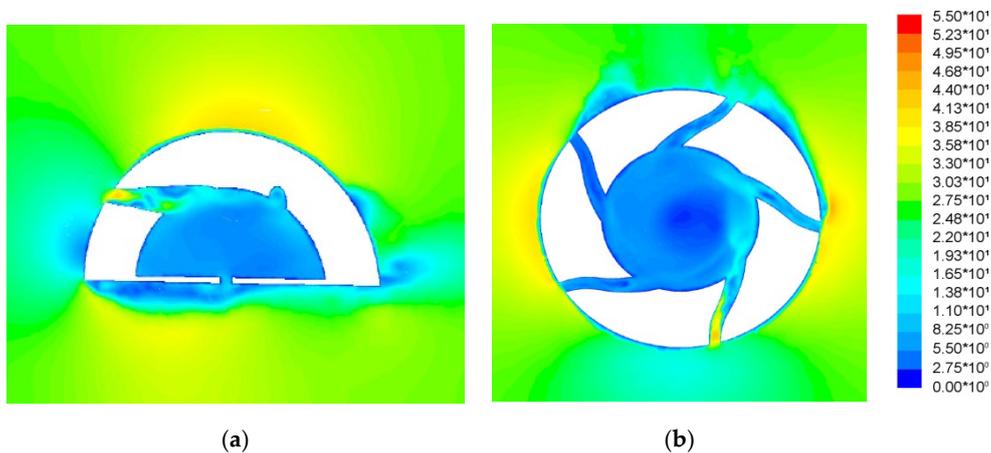

(**a**) (**b**)

**Figure 6.** The contours of velocity (m/s) of Model 2: (**a**) on a longitudinal cutting to the diffusers; (**b**) on a transverse plane in the diffusers.

From these results, it becomes evident that the area of highest pressure is located in the lower part of the cover due to the flow stagnation. For this reason, in Models 3 and 4, it was decided that the position of the diffusers had to be changed in an attempt to exploit this flow feature (they were lowered by 20 mm). Figure 7 shows the simulation results for Model 3. Figure 7a shows that the openings are now located within the zone of high pressure or the stagnation zone. Figure 7b shows the velocity contours on a transverse plane, considering the diffuser through which the wind enters. Some flow acceleration can be observed at the entrance in the central part, unlike in Models 1 and 2 where it is produced at the top. Furthermore, an area of low velocity at the top can be appreciated, indicating that the wind flow does not adapt to an increase in the cross-sections of the diffuser. To facilitate the exit of air from the interior, in addition to the ducts at the opposite side in the dome, there are also extra outlets at its base.

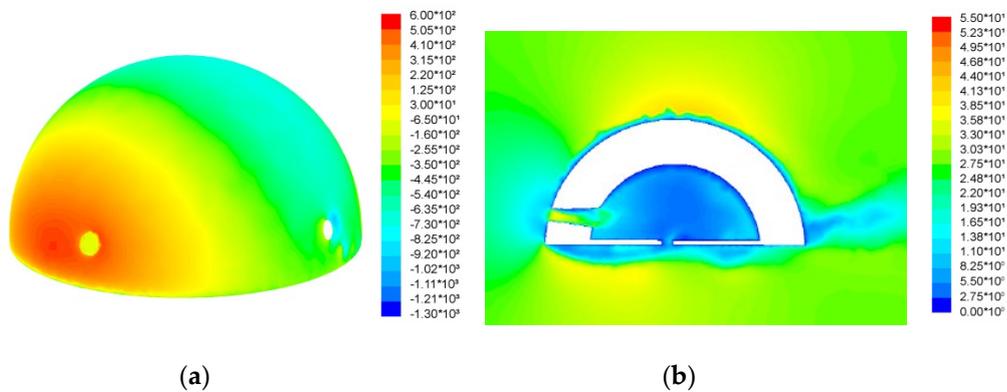

(**a**) (**b**)

**Figure 7.** Model 3: (**a**) outer cover pressure contours (Pascals); (**b**) velocity contours on a transverse plane in the diffusers (m/s).

Finally, Model 4 was proposed based on the previous results. Figure 8 shows the velocity contours for Model 4. In the previous cases, it was observed that the region of high pressure outside the dome covers a considerable part of the frontal area. Consequently, the number of diffusers was increased so that the wind enters via two openings, thereby increasing the energy that reaches the turbine. It is also worth mentioning that inside the cover, the wind revolves at a velocity near 16 m/s due to the orientation of the diffusers (Figure 8b). This was one of the purposes of this design, which is better achieved when there are more entrance openings.

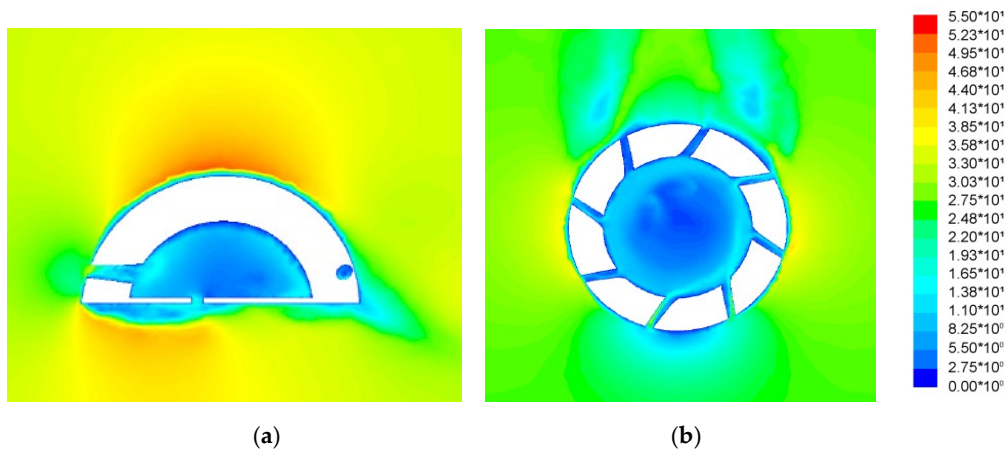

(**a**) (**b**)

**Figure 8.** The velocity contours (m/s) of Model 4: (**a**) on a longitudinal cutting to the diffusers; (**b**) on a transverse plane in the diffusers.

The above results are corroborated through the representation of velocity vectors for Models 3 and 4 in Figure 9. The latter shows the results in a transverse plane containing the diffusers, where both the inner-induced recirculating flow and the reduction of wind velocity can be observed. Furthermore, it can be noted how the inlet velocity of air into the system is higher in two of the ducts in Model 4, while in Model 3, it only occurs in one of them. This enables the force exerted on the blades to be greater for the same wind speed in the case of Model 4. On the other hand, in the upper part of Figure 9b, the exit velocity vectors are shown to be distributed over four ducts in Model 4, thus, providing more efficient air evacuation.

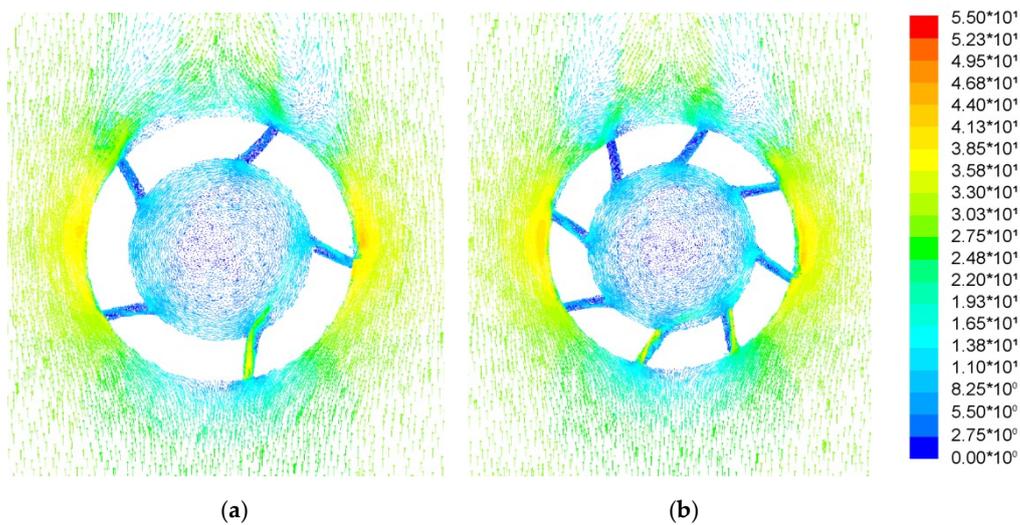

(**a**) (**b**)

**Figure 9.** Velocity vector contours on a transverse plane in the diffusers (m/s): (**a**) Model 3; (**b**) Model 4.

Figure 10 shows details of the contours of wind speed in the diffusers of the four models analyzed. The intervals of wind speed reaching the generator blades were extracted in each model. Thus, the velocity entering the inner dome (in m/s), past the discharge of the inlet ducts, in the case of Model 1 is ranged [23.5, 23.9]; in Model 2 [11.4, 13.7]; in Model 3 [17.6, 19.8]; in the case of the Model 4, it is ranged [13.0, 15.2] for the left diffuser and [17, 19.6] for the right diffuser. The latter values are related to the changes in the cross-sectional areas in the diffusers of the different models. The first model presents a smaller change between the areas at the ends of the diffuser, which means this is the model with the greatest velocity in this case. However, in a standard wind turbine, the optimal operating speed is about 15 m/s and, therefore, in a case of strong winds, as with the situation being considered herein (27 m/s), the model that delivers the best rotor speed is Model 4.

The final design in each location would be preceded by a wind speed characterization and proper configuration of the cross-section of the diffusers or nozzles to induce the desired wind speed inside. A great improvement in the device would be the inclusion of a mechanism to automatically regulate the cross-sections of the diffusers when the wind speed changes.

As for energy production, the theoretical power available in every model can be estimated. The theoretical wind power can be evaluated by

$$P = \frac{1}{2} S \rho v^3 \qquad (1)$$

where $\rho$ is the air density; $S$ is the area perpendicular to wind speed $v$.

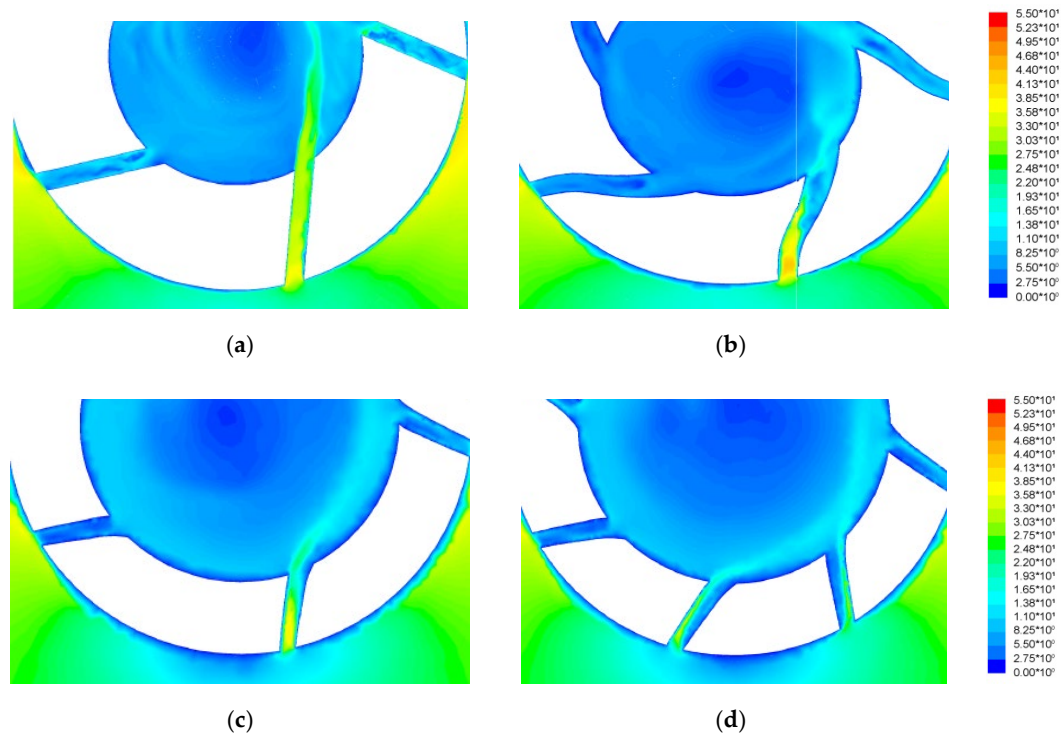

**Figure 10.** Details of the contours of the velocity of wind (m/s) of models on a transverse plane in the diffusers. (**a**) Model 1; (**b**) Model 2; (**c**) Model 3; (**d**) Model 4.

Of course, in addition to the energy available from the wind, the performance of the machinery, with its mechanical losses, and the efficiency of the electric generator have to be taken into account [8–10].

Table 4 summarizes the calculations made using the numerical results. In the case of Model 4, it has to be taken into account that the total useful power is the amount provided by the two diffusers through which the wind enters. For this reason, this is the model that provides the highest level of effective power (15,462 + 35,103 = 50,566 w); moreover, it also presents the speeds that are closest to the optimum value of a wind turbine.

**Table 4.** Areas of the diffusers and inlet air velocities and power in the models examined.

| MODEL | Output Area (mm²) | Output Area Scaled Up (mm²) | Calculated Speed (m/s) | Available Power (W) | Useful Power (W) |
|---|---|---|---|---|---|
| 1 | 43.48 | 9782.08 | 23.7 | 79.759 | 47.257 |
| 2 | 180.69 | 40,655.17 | 12.5 | 48.635 | 28.816 |
| 3 | 69.01 | 15,527.42 | 18.6 | 61.199 | 36.260 |
| 4 | 69.01 | 15,527.42 | 14.0 | 26.097 | 15.462 |
|   |       |           | 18.4 | 59.246 | 35.103 |

In Figure 11, the power available as a function of the device and outer wind speed, in the case of Model 4, is shown. The solid symbol line corresponds to the power that can be obtained from a normal wind turbine, i.e., without a cover, with a blade area equal to that of Model 4. As can be seen, when the wind speed exceeds 15–16 m/s, the turbine without a cover must stop the generator for security reasons and, thus, the power drops to zero. The open symbol line represents the power obtained with Model 4 of the patented device. The design of the diffusers in this model decreases the wind speed inside by 40%. Thus, the wind speed inside the dome reaches 15 m/s when the external wind speed is 22–23 m/s. Therefore, Model 4 allows us to keep operating the device until the external speed is about 25 m/s, namely, strong wind about 90 km/h.

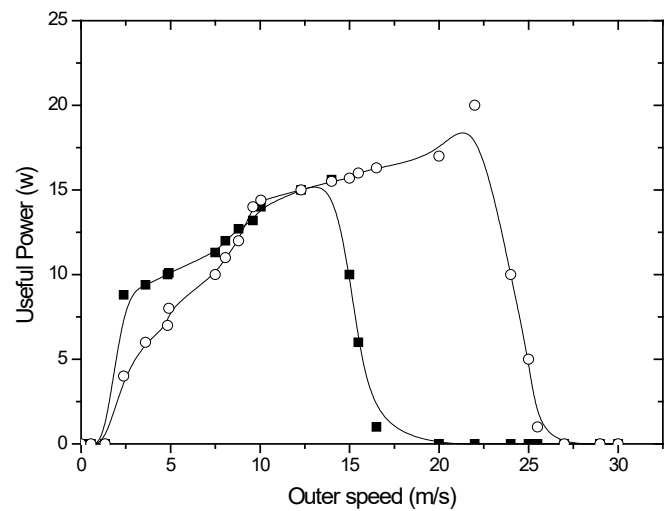

**Figure 11.** Available power in Model 4 for each outer wind speed. The solid symbol line shows the data of a normal operating system generator, and the open symbol line shows the available power in the patented device with wind speed inside the diffusers.

Lately, there have been new wind turbines with an operating range of up to 25 m/s to improve efficiency. In the model of the present work, this improvement would allow the maintenance of the operation of the wind turbine up to more than 30 m/s, thus, making possible the generation of energy in regions that present exceptionally high wind velocities.

With the data reported herein, it can be stated that with the current patented device, a wind turbine would be able to operate with prevailing wind speeds higher than those of current generators. With the dome size presented in this work, the available power is small, so its use would be ideal for small installations in places of extreme climates. However, as the device is scalable to larger sizes, far superior power could be achieved, e.g., with a dome diameter of 10 m; power up to several kilowatts could be generated.

*4.2. Experimental Results*

To complement the numerical simulations, experimental measurements on the movement of the blades were carried out. These measurements only constitute a rough first approximation to explore the best blade configuration. Two types of blades were constructed: one had a flat profile (Figure 12b), and the other a concave profile (Figure 12c). All of them had a curved shape, i.e., variable vertical frontal width, to follow the shape of the interior of the small dome (Figure 12a).

In the following sections, the results of the experiments conducted with controlled outside wind are presented. First, the axis rotation speed of the wind turbine was measured for a different number of blades, namely, 3, 4, 6, and 8 blades, and for two external speeds of wind, i.e., 3 (Figure 13a) and 6 m/s (Figure 13b). As can be seen, the rotation speed increases with the number of shaft blades, the maximum corresponding to the configuration implementing eight blades. Moreover, it is also observed, as expected, that significantly higher rotation speeds were obtained when the concave profiles are used, specifically, 10% higher speed with concave profile blades than with the flat ones. Therefore, the best configuration of the blades is that with eight blades of a concave profile.

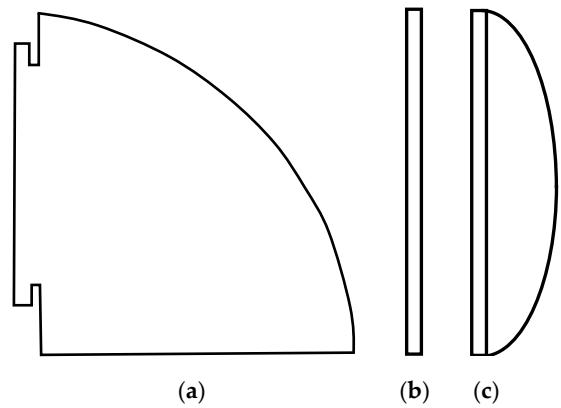

**Figure 12.** Designs of the blades used in the test of the patented model. (**a**) Front view of the two blades; (**b**) flat profile of the first design; (**c**) concave profile of the second design.

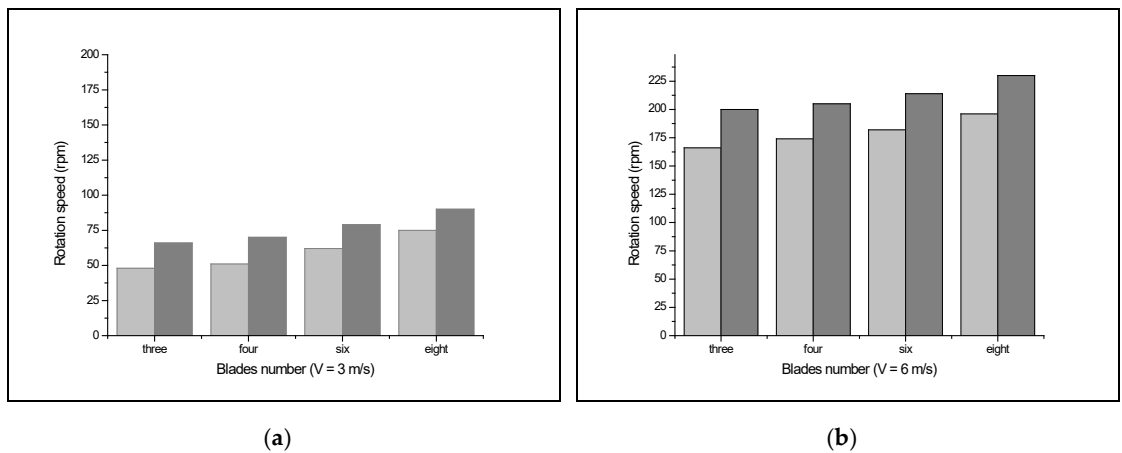

**Figure 13.** Rotation speed of the wind turbine as a function of the number of blades (3, 4, 6, 8). (**a**) External wind speed 3 m/s; (**b**) 6 m/s. Light grey bars correspond to flat profile blades; dark grey bars to concave profile blades.

Finally, once the best blade configuration was established, the study was extended to analyze the power curve. Figure 14 shows how the rotation speed of the wind turbine increases with the increase of outer wind speed. This increase reaches a maximum and then stabilizes around 300 rpm for external wind speed greater than 15 m/s. This result coincides with that obtained in Figure 11, and the power is kept at high values up to the maximum speed tested. The tip speed ratio (TSR) is the ratio between blade peripheral speed and the wind speed. The dependency between TSR of the blades and wind power is represented in Figure 15. Furthermore, this value is shown to increase quickly until reaching a maximum for low wind power, to subsequently decrease with a smoother slope. This would indicate that the efficiency of the turbine considered will be better for non-high wind speeds.

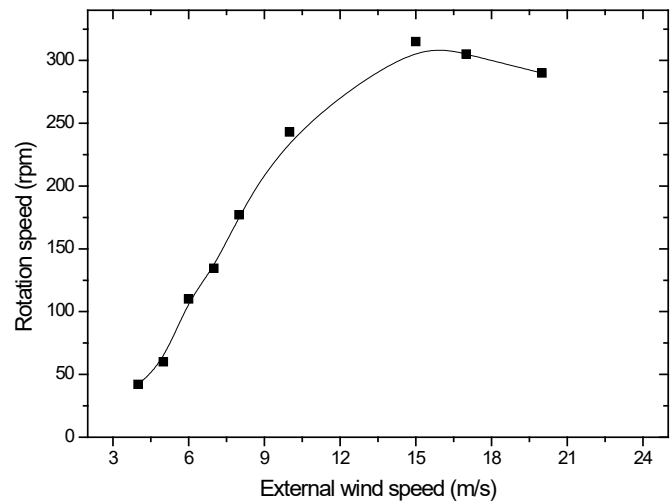

**Figure 14.** Rotation speed of the wind turbine (rpm) as a function of outer wind speed (m/s).

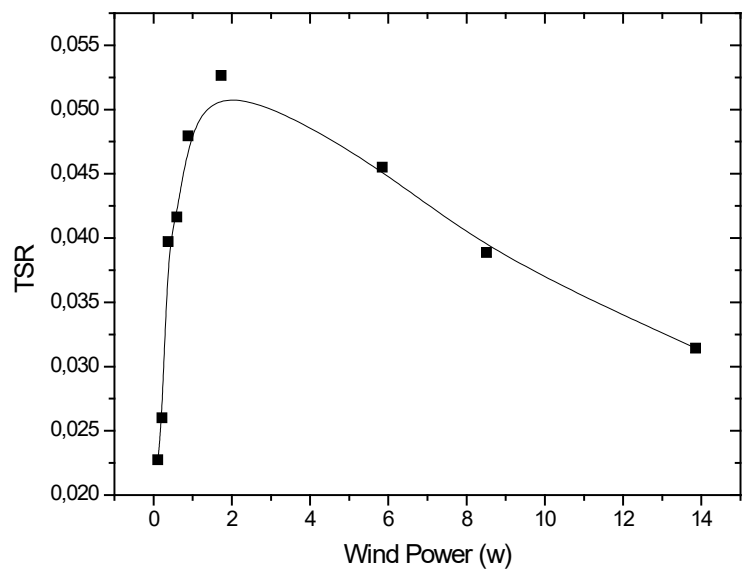

**Figure 15.** Tip speed ratio (TSR) of the wind turbine as a function of wind power (W).

## 5. Conclusions

In the present paper, the design and operation of a patented device to cover VAWTs and improve their operation has been proposed and assessed. According to the text of the patent, this system is meant to allow the use of wind power installations in places with special weather conditions, where, so far, it is difficult or unprofitable. The variable geometry of the ducts that direct the wind to the turbine blades can both enhance and reduce the wind speed entering the inner dome. In this last case, the device is suitable for installations in low wind areas and, especially as the most important innovation, in places where there are very strong winds.

In locations where there are very strong, maintained, or sporadic winds (hurricanes or typhoons), towers of normal horizontal-axis turbines installed may suffer serious damage. The cover proposed in the patent protects the structure so that it does not fall, and external openings can be closed with automatic doors if necessary (rain, snow, or excessive speeds). Therefore, this design can be used to generate renewable electricity of low or middle power in remote locations (such as facilities close to the polar circles) or high power in the areas of hurricanes.

In the simulations performed on the initial designs, different options to improve the system have been proposed. The areas of greatest pressure are found in a region located lower in height than that of the openings of the ducts in the initial model, covering a significant part of the frontal area. Therefore, the height of the openings was lowered and the number of ducts increased to eight openings in the final design, Model 4. The results for this model showed optimum values of wind speeds inside the cover and a considerable increase in the power that can be generated.

Specifically, the optimum operating speed of the turbine is obtained inside the dome (<20 m/s), with wind speeds on the outside near to 100 km/h. Therefore, it would be possible to produce energy with a wind turbine in places with strong prevailing winds without stopping power generation. Any new improvement in modern wind turbines can be applied to our model, for example, operation at higher speeds.

Finally, further experiments to determine the best configuration of the blades were conducted. For different wind speeds, within the operating range of the wind turbine, the best effectiveness in axis movement corresponded to the configuration with eight blades of a concave profile. Therefore, the device has been shown to be suitable for use in regions with extreme weather conditions.

**Author Contributions:** Conceptualization, M.B.J.A.; methodology, M.B.J.A. and G.M.C.; software, E.G.A.J. and G.M.C.; formal analysis, M.B.J.A. and G.M.C.; investigation, M.B.J.A. and G.M.C.; resources, M.B.J.A.; data curation, M.B.J.A., G.M.C., and E.G.A.J.; writing—original draft preparation, M.B.J.A.; writing—review and editing, M.B.J.A. and G.M.C.; supervision, M.B.J.A. All authors have read and agreed to the published version of the manuscript.

**Funding**: This research received no external funding.

**Acknowledgments:** We acknowledgment the University of Jaén for the resources used.

**Conflicts of Interest:** The authors declare no conflict of interest.

*Sustainability* **2020**, *12*, x FOR PEER REVIEW 16 of 1710. Barnes, R.H.; Morozov, E.V.; Shankar, K. Improved methodology for design of low wind speed specific wind turbine blades. *Compos. Struct.* **2015**, *119*, 677–684.
11. Lecuona Neumann, A. La Energía Eólica: Principios Básicos y Tecnología. Book: Universidad Carlos III, Madrid, Spain, 2002.
12. Armada, M.J. Energía Eólica. Master's Thesis, Master Europeo en Energías Renovables y Eficiencia Energética. Universidad de Zaragoza, Zaragoza, Spain, 2001.
13. Manwell, J.F.; McGowan, J.G.; Rogers, A.L. *Wind Energy Explained: Theory, Design and Application*, 2nd ed.; John Wiley & Sons Ltd.: West, Sussex, UK, 2002.
14. Damota, J.; Lamas, I.; Couce, A.; Rodriguez, J. Vertical Axis Wind Turbines: Current Technologies and Future Trends. In Proceedings of the International Conference on Renewable Energies and Power Quality, La Coruna, Spain, 25–27 March 2015.
15. Riegler, H. HAWT versus VAWT: Small VAWTs find a clear niche. *Refocus* **2003**, *4*, 44–46.
16. Mertens, S. Wind energy in urban areas: Concentrator effects for wind turbines close to buildings. *Refocus* **2002**, *3*, 22–24.
17. Mahmoud, N.H.; El-Haroun, A.A.; Wahba, E.; Nasef, M.H. An experimental study on improvement of Savonius rotor performance. *Alexandria Eng. J.* **2012**, *51*, 19–25.
18. Gupta, R.; Biswas, A.; Sharma, K.K. Comparative study of a three-bucket Savonius rotor with a combined three-bucket Savonius—Three-bladed Darrieus rotor. *Renew. Energy* **2008**, *33*, 1974–1981.
19. Sukanta, R.; Ujjwal, K.S. Wind tunnel experiments of a newly developed two-bladed Savonius-style wind turbine. *Appl. Energy* **2015**, *137*, 117–125.
20. Elkhoury, M.; Kiwata, T.; Aoun, E. Experimental and numerical investigation of a three-dimensional vertical-axis wind turbine with variable-pitch. *J. Wind Eng. Ind. Aerodyn.* **2015**, *139*, 111–123.
21. Burlando, M.; Ricci, A.; Freda, A.; Repetto, M.P. Numerical and experimental methods to investigate the behaviour of vertical-axis wind turbines with stators. *J. Wind Eng. Ind. Aerodyn.* **2015**, *144*, 125–133.
22. Roy, S.; Saha, U.K. Review on the numerical investigations into the design and development of Savonius wind rotors. *Renew. Sustain. Energy Rev.* **2013**, *24*, 73–83.
23. Roy, S.; Saha, U.K. Review of experimental investigations into the design, performance and optimization of the Savonius rotor. *SAGE J.* **2013**, *227*, 528–542.
24. Danao, L.A.; Eboibi, O.; Howell, R. An experimental investigation into the influence of unsteady wind on the performance of a vertical axis wind turbine. *Appl. Energy* **2013**, *107*, 403–411.
25. Maeda, T. Development of simplified design method of vertical axis wind turbine. *Jpn. Wind Energy Assoc. Wind Energy* **2012**, *36*, 360–363.
26. Bhutta, M.M.A.; Hayat, N.; Farooq, Ahmed, U.; Ali, Z.; Jamil, S.R.; Hussain, Z. Vertical axis wind turbine e a review of various configurations and design techniques. *Renew. Sustain. Energy Rev.* **2012**, *16*, 1926–1939.
27. Aoki, S.; Kogaki, T. Design elements of horizontal axis wind turbine. *Jpn. Wind Energy Assoc. Wind Energy* **2012**, *36*, 355–359.
28. Pourrajabian, A.; Afshar, P.A.N.; Ahmadizadeh, M.; Wood, D. Aero-structural design and optimization of a small wind turbine blade. *Renew. Energy* **2016**, *87*, 837–848, doi:10.1016/j.renene.2015.09.002.
29. Li, Q.; Maeda, T.; Kamada, Y.; Murata, J.; Furukawa, K.; Yamamoto, M. Effect of number of blades on aerodynamic forces on a straight-bladed Vertical Axis Wind Turbine. *Energy* **2015**, *90*, 784–795, doi:10.1016/j.energy.2015.07.115.
30. Kim, D.; Gharib, M. Efficiency improvement of straight-bladed vertical-axis wind turbines with an upstream deflector. *J. Wind. Eng. Ind. Aerodyn.* **2013**, *115*, 48–52, doi:10.1016/j.jweia.2013.01.009.
31. Marsh, P.; Ranmuthugala, D.; Penesis, I.; Thomas, G. Three-dimensional numerical simulations of straight-bladed vertical axis tidal turbines investigating power output, torque ripple and mounting forces. *Renew. Energy* **2015**, *83*, 67–77, doi:10.1016/j.renene.2015.04.014.
32. Goude, A.; Ågren, O. Simulations of a vertical axis turbine in a channel. *Renew. Energy* **2014**, *63*, 477–485, doi:10.1016/j.renene.2013.09.038.
33. Almohammadi, K.; Ingham, D.; Ma, L.; Pourkashanian, M. Modeling dynamic stall of a straight blade vertical axis wind turbine. *J. Fluids Struct.* **2015**, *57*, 144–158, doi:10.1016/j.jfluidstructs.2015.06.003.
34. Ying, P.; Chen, Y.; Xu, Y. An aerodynamic analysis of a novel small wind turbine based on impulse turbine principles. *Renew. Energy* **2015**, *75*, 37–43, doi:10.1016/j.renene.2014.09.035.
35. Allaei, D.; Tarnowski, D.; Andreopoulos, Y. INVELOX with multiple wind turbine generator systems. *Energy* **2015**, *93*, 1030–1040, doi:10.1016/j.energy.2015.09.076.